\documentstyle[aps,twocolumn,psfig,epsfig,floats,citesort]{revtex}

\def\eps{{\epsilon}}

\def\begineq{\begin{equation}}
\def\endeq{\end{equation}}
\def\be{\begin{equation}}
\def\ee{\end{equation}}

\begin{document}
\bibliographystyle{prsty}

\title{
On geometry effects in Rayleigh-B\'enard convection
}
\author{Siegfried Grossmann$^1$ and Detlef Lohse$^2$}
\address{
$^1$ Department of Physics, University of Marburg, Renthof 6, 
D-35032 Marburg, Germany\\
$^2$Department of Applied Physics,
University of Twente, 7500 AE Enschede, Netherlands\\
}

\date{\today}

\maketitle

\begin{abstract}
Various recent experiments hint at a geometry dependence 
of scaling relations in Rayleigh-B\'enard convection. 
Aspect ratio and shape dependences have been found. 
In this paper a mechanism is offered which can account 
for such dependences.
It is based on Prandtl's theory for laminar boundary layers
and on the conservation of volume flux of the large scale wind. 
The mechanism implies the possibility of different 
thicknesses of the kinetic boundary layers at the sidewalls and the 
top/bottom plates, just as experimentally found by Qiu and Xia
(Phys. Rev. E58, 486 (1998)), and also different $Ra$-scaling of the 
wind measured over the plates and at the sidewalls. 
In the second part of the paper a scaling argument for the 
velocity and temperature fluctuations in the bulk is developed. 
\end{abstract}

\vspace{0.8cm}

Turbulent Rayleigh-B\'enard convection is one of the classical problems in
fluid dynamics \cite{kad01,ahl02,sig94}. The great interest into this problem
presumably also originates from the relevance of thermal turbulence in
meteorology, geophysics, oceanography, and astrophysics. However, in contrast
to these natural problems, in laboratory experiments the thermal convection 
is confined to a container. Recent experiments have revealed that geometrical 
details of this container are important. 
Daya and Ecke \cite{day01} find that the bulk
temperature and velocity fluctuations
in a cylindrical cell and in a square cell are very different from each
other. 
For cylindrical cells the aspect ratio dependence 
has been examined by the Ahlers group \cite{xu00,ahl01}
and by the Tong group \cite{qiu01b}; in those experiments only the
prefactors of effective power laws seem to change.  
Niemela and Sreenivasan \cite{nie02} find an aspect ratio dependence
of the heat flow at large Rayleigh numbers, and conclude
that the focus of the next generation of experiments must be on large aspect
ratio cells. 

Another recent surprising result to be mentioned
in this context is the dependence of the 
kinetic boundary layer (BL) thickness 
(which is set by the mean flow velocity profile above the wall)
on its location in the cell.
Qiu and Xia \cite{qiu98,lam02} find that it scales differently at the
sidewalls and at the top or bottom plates. 
They find $\delta_w/ L = 
3.6~ Ra^{-0.26 \pm 0.03}$ for the sidewall BL thickness 
(from now on abbreviated with {\it w}, for wall) and 
$\delta_p /L =0.65 Ra^{-0.16 \pm 0.02} Pr^{0.24\pm 0.01}$
for the thickness at the top or bottom plates
(from now on abbreviated with {\it p}, for plate), both for
an aspect ratio $\Gamma= 1$ cylindrical cell.
No existing theory of Rayleigh-B\'enard convection accounts for 
this difference. Our own unifying theory of thermal convection
\cite{gro00,gro02} so far allows only for {\it one} kinetic boundary
layer thickness, and gives a power law exponent $-0.22$ (for
$\Gamma = 1$ and a Prandtl number of $Pr=6$) \cite{gro02}, close
to the average $(-0.16 - 0.26)/ 2 = -0.21$ of the experimental 
results for the plate and the wall BLs thicknesses.

In this paper we set out to offer a mechanism 
to account for the shape and aspect ratio dependence in Rayleigh-B\'enard
convection, and for the difference between the wall and plate kinetic BLs. 
The starting point is Prandtl's BL theory \cite{pra04,bla08,ll87}. 
The main
physical ingredient is the conservation of the volume flux of the 
large scale wind.
In the second part of the paper we 
present scaling arguments for the fluctuations
of the  velocity and the  temperature in the bulk. 

\vspace{0.2cm}

Laminar BL flow is described by
the famous  Prandtl equations \cite{pra04,bla08,ll87}. 
In these equations streamwise lengths are 
scaled by the streamwise length scale $l$, while wall-normal lengths are
(re)scaled by $l/ \sqrt{Re}$. Here, $Re= l U_0/\nu$ is the Reynolds number
based on the streamwise length scale $l$ and the streamwise velocity scale
$U_0$. Correspondingly, streamwise velocities are scaled by 
$U_0$ and wall-normal velocities by $U_0/\sqrt{Re}$. 
The immediate consequence is that the thickness of the laminar BL 
scales as $\delta \sim l/\sqrt{Re}$
(\ \cite{pra04}; see also e.g. section 39 of ref.\ \cite{ll87}). 

In the context of Rayleigh-B\'enard convection this means that
the relevant streamwise length scales 
are the width $d$ of the cell for the plates and the height $L$ of the 
cell for the walls, i.e., 
{\it different} for cells with non-unity aspect ratio $\Gamma = d/L$. 
Correspondingly, according to the Prandtl theory, also the widths
of the BLs at the walls and at the plates are different, namely,
\be
\delta_w = a {L \over \sqrt{U_w L/\nu }} 
\label{deltaw}
\ee
and 
\be
\delta_p = a {d \over \sqrt{U_p d/\nu }},
\label{deltap}
\ee
respectively.
Here, $a$ is a dimensionless prefactor of order 1; Blasius \ \cite{bla08} 
gave $a=1.72$ for a semi-infinite kinetic BL, cf. also \cite{ll87}. 
If we assumed that the streamwise wall and plate velocities are the
same, $U_w = U_p$, we would immediately get aspect ratio dependent BL widths 
$\delta_w = \delta_p / \sqrt{\Gamma }$.

\begin{figure}[htb]
\setlength{\unitlength}{1.0cm}
\begin{picture}(11,9)
\put(2.0,8.7){\large(a)}
\put(7.0,8.7){\large(b)}
\put(1.0,0.6)
{\epsfig{figure=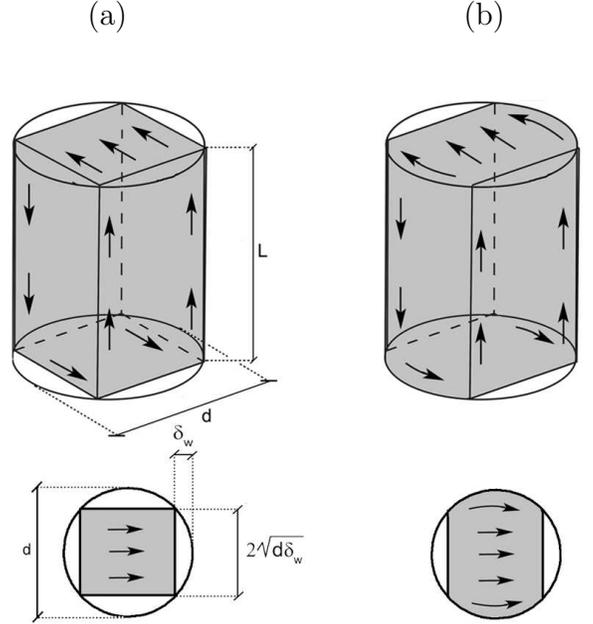,width=10cm,angle=-0}}
\end{picture}
\caption[]{
Confined (a) and unconfined (b) flow in a cylindrical cell. The upper 
drawings show
sketches of the 3D flow, the lower ones a top view on the bottom plate.
The flow area above the plate is shaded. The spanwise extension 
of the wind
along the walls 
is $2 \sqrt{d\delta_w}$ and follows from trigonometry.  
}
\label{fig1}
\end{figure}

However, the experiments by Tong's group \cite{qiu01b} show that the
assumption $U_w = U_p$ does not hold. Indeed, a more realistic assumption 
seems to us {\it volume flux conservation}, i.e., equal volume fluxes of the 
wind over the walls and the plates, 
\be
\dot V_w = \dot V_p \ .
\label{volflux}
\ee  
Then the geometry of the cell immediately enters. 
We will discuss two extreme situations: The ``confined'' 
flow, and the ``unconfined'' flow, see figure \ref{fig1}.
In the confined flow the convection roll has the  {\it same}
spanwise extension both at the plates and the walls. In contrast, 
the unconfined flow makes use of the full plate width $d$ in 
the middle, whereas towards the walls its extension is of order 
$2\sqrt{d \delta_w}$ \cite{bem1}. 
The real flow will presumably be in between these two extremes.

{\it Confined flow:} 
The spanwise length scale of the flow is of order
$2\sqrt{d \delta_w}$ at the walls and at the plates 
(see figure \ref{fig1}).
The corresponding volume fluxes are 
$\dot V_w = 2\sqrt{d \delta_w} \delta_w U_w$ at the walls and
$\dot V_p = 2\sqrt{d \delta_w} \delta_p U_p$ at the plates.
With volume flux conservation eq.\ (\ref{volflux}) and the 
Prandtl relations (\ref{deltaw}) and (\ref{deltap}) we 
immediately get 
\be
{\delta_w / \delta_p} =  \Gamma^{-1} \ ,
\label{confined_del}
\ee
\be
{U_p /  U_w } = \Gamma^{-1} \ .
\label{confined_u}
\ee
Thus the wall
and plate BLs have different thicknesses. But there is {\it no}  
difference in the Ra-scaling of $\delta_w$ and $\delta_p$,
in contrast to what  was found by
Xia's group \cite{qiu98}.

{\it Unconfined flow:} 
The spanwise length scale of the flow at the cylindrical walls still is
$2\sqrt{d \delta_w}$. 
But now by definition the flow at the plates is
unconfined spanwise and the respective spanwise extension is of order $d$. 
The corresponding volume fluxes are now 
$\dot V_w = 2\sqrt{d \delta_w} \delta_w U_w$ at the wall and
$\dot V_p = d \delta_p U_p$ at the plate.
With volume flux conservation eq.\ (\ref{volflux}) and the 
Prandtl relations (\ref{deltaw}) and (\ref{deltap}) we now 
obtain $U_p^2 /U_w = \nu L^{-1} \Gamma^{-4} 2^4 a^4$ and
\be
{\delta_w / \delta_p} = \Gamma^{-1/2} \sqrt{{U_p / U_w}} \ .
\ee
We now introduce wall and plate Reynolds numbers,
\be
Re_w = U_w L/\nu \ , \qquad 
Re_p = U_p L/\nu \ .
\label{re}
\ee
Note that both are defined with the height $L$ of the cell.
With these definitions we find
\be
{\delta_w / \delta_p} = \Gamma^{-5/2} 4a Re_p^{-1/2}
=\sqrt{4a} \Gamma^{-3/2} Re_w^{-1/4},
\label{unconfined_del}
\ee
\be
{U_p / U_w } = 
{Re_p / Re_w } = 
\Gamma^{-4} 16a^2 Re_p^{-1}=
4a \Gamma^{-2} Re_w^{-1/2}.
\label{unconfined_u}
\ee

Equation (\ref{unconfined_del}) is our first main result.
It shows, for unconfined flow, indeed a Reynolds 
number and thus a Rayleigh number dependence 
of the ratio between the BL thicknesses at the wall
and at the plate, due to the cylindrical shape of the container. 
$\delta_w$ and $\delta_p$ scale
differently with $Ra$, just as experimentally observed \cite{qiu98}.

Since experimental flow is expected to lie in between the confined and the 
unconfined cases, $\delta_w /\delta_p$ 
should be between $Re_p^{-1/2}$ and $Re_p^{0}$ or between 
$Ra^{-0.22}$ and $Ra^{0}$; here we have used the 
effective scaling $Re_p \sim Ra^{0.45}$ which holds for 
$Pr=5$ and $\Gamma = 1$ both experimentally \cite{qiu01b,bem2} 
and theoretically \cite{gro02} in the $Ra$ regime 
$10^8 < Ra < 10^{10}$. 
Indeed, the experimental ratio 
has a scaling exponent in between the values given by the 
unconfined and confined cases, namely
$\delta_w /\delta_p \sim Ra^{-0.11} $ \cite{qiu98}.

 \begin{table}
 \begin{center}
 \begin{tabular}{|c|c|c|c|c|}
 \hline
       &  model
       &  $\Gamma = 1$
       & $\Gamma = 1/2$
       & $\Gamma = 2$
\\
\hline
      $\delta_w /\delta_p$ 
      &  experi., extrapol: confined
      &   1
      &   2.1
      &   0.29
\\
      &  experi., extrapol: unconfined
      &   1
      &   3.3
      &   0.19
\\
      &  theory, confined
      &   1
      &   2.0
      &   0.50
\\
      &  theory, unconfined
      &   1
      &   5.6
      &   0.18
\\
\hline
      $U_p /U_w$ 
      &  experi., extrapol: confined
      &   1
      &   1.2
      &   2.0
\\
      &  experi., extrapol: unconfined
      &   1
      &   3.1
      &   0.8
\\
      &  theory, confined
      &   1
      &   2.0
      &   0.5
\\
      &  theory, unconfined
      &   1
      &   16
      &   0.063
\\
 \hline
 \end{tabular}
 \end{center}
\caption[]{
$\delta_w/\delta_p$ and $U_p/U_w = Re_p/Re_w$ for
different aspect ratios, gauged to $\Gamma = 1$, where we have put these
ratios to 1. The Reynolds number $Re_p=3.7 \cdot 10^9$
is fixed and the experimental values are
extrapolated to that $Re_p$, see text.  
Note that the error in reading off in particular
the boundary layer widths from 
the figures 3,6, and 8 of ref.\ \cite{qiu01b} is at least 25\%
and therefore the experimental ratios
should only be taken as a very rough estimate. 
}
\label{tab1}
 \end{table}

We now come to our other result, eq.(\ref{unconfined_u}), and to the 
aspect ratio dependence of $\delta_w/\delta_p$ and $U_p/U_w$. 
For $\Gamma = 1/2$ and $\Gamma = 2$ these ratios
(as predicted in the two cases, and gauged to $\Gamma = 1$) are 
summarized in table \ref{tab1}. The experimental data are obtained from
figure 3 ($\Gamma =1$, $Ra=3.7 \cdot 10^9$), 
figure 6 ($\Gamma =2$, $Ra=4.9 \cdot 10^8$), and  
figure 8 ($\Gamma =1/2$, $Ra=3.28 \cdot 10^{10}$) 
of reference \cite{qiu01b}. In this series of  
figures both $\Gamma$ and $Ra$ were changed at the same time.  
In order to compare the ratios at $\Gamma = 2$ measured 
at $Ra=5\cdot 10^8$ with those at $\Gamma = 1$ measured
at $Ra= 3.7 \cdot 10^9$, they first have to be extrapolated
to this higher $Ra$ number. Correspondingly, in the $\Gamma=1/2$
case we have to extrapolate the ratios from 
the original value $Ra=3.3 \cdot 10^{10}$ to $Ra= 3.7 \cdot 10^9$.
These extrapolations have been done according to the unconfined case
formulas (\ref{unconfined_del}), (\ref{unconfined_u}) or
according to the confined case ones 
(\ref{confined_del}), (\ref{confined_u}).
In the latter case in fact no extrapolation is necessary
as in those equations there is no $Ra$ or $Re$ dependence. 

From table \ref{tab1} we conclude that  
the experimental results (if extrapolated with the unconfined 
flow model)
in general lie in between the predictions of the confined and unconfined cases,
just as expected. An exception is $U_p/U_w$ for $\Gamma = 2$; we have 
no explanation for that. If extrapolated with the confined flow model 
both velocity ratios
do not lie in between the predictions of the two flow cases.


The physics of the aspect ratio dependence is as follows: 
For increasing aspect ratio the available plate area of
the flow increases and the plate velocity can go down. 
According to the Prandtl law (\ref{deltap}), the plate BL thickness then
increases. At the walls there is no additional lateral space for 
larger aspect ratios, therefore the flow has to accelerate,
$U_w > U_p$, and correspondingly $\delta_w < \delta_p$. 

Note that above  dependence on $\Gamma$ was derived assuming 
there is {\it one} convection roll. Once the
convection develops two rolls on top of each other (for $\Gamma$ less than 1) 
or next to each other (for $\Gamma$ larger than 1), the aspect
ratio dependence has to be embodied in a more sophisticated way.
For small $\Gamma  \le 1/2$  a transition from one roll to two
rolls has been observed numerically \cite{wer93,ver02}.
Possibly also in experiment both states of convection
have been realized, see the discussion in \cite{nie02}. 
Thus one cannot formally consider 
$\Gamma \to 0$ or $\Gamma \to \infty$ in the relations 
(\ref{confined_del}) -- (\ref{unconfined_u}).

Can one manipulate a system such that it is no longer in between 
the confined and the unconfined cases? An option may be to take 
a {\it square box} and slightly {\it tilt} it in order to disfavor the 
flow along the diagonal, which would
occur without tilting \cite{day01}.
 If this can be achieved, one expects 
such flow to follow the confined flow model with confinement by
the left and right walls. 
For this type of flow eq.\ (\ref{confined_del}) 
should hold, i.e., the wall and the plate BLs should show the 
same $Ra$-scaling.
An experimental test of this predictions seems worthwhile
to us to check the proposed theory.


{\it Kinetic and thermal dissipation in the BLs:}
Within above framework 
the aspect ratio dependence can also be embodied into 
the unifying theory of thermal convection \cite{gro00,gro02}.
The main idea of that theory is to split the kinetic dissipation
$\eps_u$ and the thermal dissipation $\eps_\theta$, for
which exact global relations can be derived from the
Boussinesq equations, into their bulk and BL contributions.


For the kinetic dissipation this splitting obviously has to
be extended to 
\be
\eps_u = 
\eps_{u,p}+
\eps_{u,w}+
\eps_{u,b} \ ,
\label{split_eps_u}
\ee
where 
$\eps_{u,p}$, $\eps_{u,w}$ mean the kinetic dissipation in the plates' or 
walls' BLs, and $\eps_{u,b}$ the bulk kinetic dissipation.
Following \cite{gro00}, the BL dissipations are estimated as
\be
\eps_{u,p} = \nu {U_p^2 \over \delta_p^2 } {\delta_p \over L}
= {\nu^3 \over L^4} {1\over a \Gamma^{1/2}} Re_p^{5/2}={\nu^3 \over L^4}
2^5 a^{3/2} \Gamma^{-9/2} Re_w^{5/4}
\label{eps_u_p}
\ee
and 
\be
\eps_{u,w} = \nu {U_w^2 \over \delta_w^2 } {\delta_w \over l}
= {\nu^3 \over L^4} {\Gamma^9 \over 2^{10} a^6 } Re_p^{5}
={\nu^3 \over L^4} {1 \over a \Gamma} Re_w^{5/2}.
\label{eps_u_w}
\ee
The bulk dissipation will be dealt with below.

For the thermal dissipation the partition into BLs and bulk is  
different. Though at least in ideal RB convection the sidewalls are perfectly
isolated (no flux condition), thermal BLs can develop at the sidewalls,
as observed both in numerical simulations and in experiment
\cite{ahl00,chi01a,ver02,qiu01b}. However, within our unifying theory
ref.\ \cite{gro00,gro02} it is not necessary to distinguish
between sidewall thermal dissipation and bulk
thermal dissipation. Therefore, here
we introduce the notation $\eps_{\theta,p}$ for the thermal dissipation
in the plate BLs and  $\eps_{\theta,\bar p}$ for the thermal 
dissipation elsewhere, i.e.,
within the bulk {\it and} within the sidewall BLs (``non-plate'').

To estimate these contributions we remind that $\eps_{\theta} = \kappa \Delta^2
L^{-2} Nu$ and that the heat current $Nu$ consists of two terms
\begin{eqnarray}
Nu &=&  \left( \left< u_z \theta\right>_{A,t}(z) 
-\kappa \partial_z \left< \theta \right>_{A,t}(z) \right) / (\kappa 
{\Delta L^{-1}}) \nonumber \\ &=& 
\left( \eps_{\theta, \bar p} + \eps_{\theta,  p} \right) / 
(\kappa \Delta^2 L^{-2} )
= \eps_\theta / (\kappa \Delta^2 L^{-2} )
. 
\label{balance}
\end{eqnarray}
Here $\left< ... \right>_{A,t}$ denotes the average on time and $x,y$-plane. 

For $z$ {\it outside} of the plates' thermal BLs 
the first term in eq.\ (\ref{balance}) dominates.
We estimate $\left< u_z \theta\right>_{A,t} \sim U \Delta$
and obtain $\eps_{\theta, \bar p}/(\kappa \Delta^2/L^2)
 \sim Pr Re$. Note again that the 
thermal dissipation in the sidewall BLs is included in 
$\eps_{\theta, \bar p}$ and may even be
dominant as compared to the contributions from the bulk, 
as the numerical simulations by Verzicco 
and Camussi suggest \cite{ver02}.

For $z$ {\it within} the plates' thermal BLs, i.e., $z/L \approx 0$ or 
$z/L \approx 1$, the second term in eq.\ (\ref{balance}) dominates.
Estimating the temperature gradient as $\Delta /\lambda_\theta$ gives
$\eps_{\theta , p}/(\kappa \Delta^2/L^2)
 \sim L / \lambda_\theta$. To connect the BL width 
$\lambda_{\theta}$ with $Re$, we
use the temperature equation 
$
\partial_t \theta = u_j \partial_j \theta + 
\kappa \partial_j\partial_j \theta $,
leading to $U d^{-1} \sim \kappa \lambda_{\theta}^{-2}$, which implies 
$\eps_{\theta, p} / (\kappa \Delta^2/L^2)
\sim \sqrt{Re Pr}$, cf.\ \cite{gro00}. According to this
derivation the Reynolds number one has to use in $\eps_{\theta, \bar p}$ 
is $Re_w$, while $\eps_{\theta, p}$ is determined by $Re_p$.

{\it Scaling relations for bulk velocity and temperature fluctuations:}
Up to now we 
dealt with the BLs. As for the bulk fluctuations $u'$, 
the experiments \cite{lam02} suggest that they scale differently from
the large scale velocity $U$, namely with a weaker
$Ra$ dependence. Typically, the fluctuation Reynolds number is
$Re'=u'L/\nu \sim Ra^{0.40\pm 0.03}$  and the
large scale velocity Reynolds number with $Re= UL/\nu 
\sim Ra^{0.43 ... 0.50} $ \cite{cas89,nie01,lam02,sig94,qiu01b}. 
In refs.\ \cite{gro00,gro02}
we developed a theory for the dependences $Re(Ra,Pr)$ and
$Nu(Ra,Pr)$. Here we add how the fluctuations $Re'$ 
and $\theta'$ should behave as functions of $Ra$ and $Pr$. 

The turbulence in the bulk is driven by the large scale wind, therefore
$\eps_{u,b} \sim U^3/L$ as used in refs.\ \cite{gro00,gro02}.
The energy cascades down and will be dissipated on scales comparable
to the Kolmogorov scale $\eta = (\nu^3 / \eps_{u,b})^{1/4}$. Thus we can 
estimate the bulk dissipation also with 
$\eps_{u,b} \sim \nu u^{\prime 2}/\eta^2$. Combining both we obtain
\be
Re'\sim Re^{3/4} \ . 
\label{reprime}
\ee
With $Re\sim Ra^{0.43...0.50}$ this means 
$Re'\sim Ra^{0.32...0.38}$, which is close to above quoted experimental
finding $Re'\sim Ra^{0.40\pm 0.03}$ by Lam et al. \cite{lam02}
and also to Daya and Ecke's finding 
\cite{day01} $Re' \sim Ra^{0.36\pm 0.05}$ for a 
square cell.

The same reasoning can be followed to calculate the scaling
of the typical bulk temperature fluctuations $\theta'$. 
The thermal bulk dissipation is driven by the large scale 
temperature $\Delta$; its time scale again is $L/U$.
Thus $\eps_{\theta, b} \sim \Delta^2 U/L$. On the other hand,
the thermal energy in the bulk is
dissipated by the temperature fluctuations $\theta'$ on scale 
$\eta_{\theta} = (\kappa^3/\eps_{u,b})^{1/4}\sim L Pr^{-3/4} Re^{-3/4}$,
i.e., $\eps_{\theta, b} \sim \kappa \theta^{\prime 2} /\eta_\theta^2$.
Balancing these two expressions for $\eps_{\theta, b}$ one obtains
\be
{\theta' / \Delta} \sim Pr^{-1/4} Re^{-1/4} \ . 
\label{thetaprime}
\ee
With the experimental result \cite{lam02} 
$Re\sim Ra^{0.43} Pr^{-0.76}$, which holds in
$10^8 < Ra < 3\cdot 10^{10}$ and
$3< Pr< 1200$, this means
${\theta' / \Delta} \sim Pr^{-0.06} Ra^{-0.11}$.
For $Re\sim Ra^{1/2}$ \cite{cas89,nie01} one gets 
${\theta' / \Delta} \sim  Ra^{-0.13}$. Both power laws are  very close
to the experimental findings ${\theta' / \Delta} \sim Ra^{-1/7}$
\cite{cas89} or  ${\theta' / \Delta} \sim Ra^{-0.10 \pm 0.02}$
\cite{day01} for cylindrical cells.

In conclusion, 
we have offered a possible mechanism to cope with the recently observed effects
of the RB cell geometry and 
the different $Ra$ behaviors of the kinetic BLs near
the plates and the sidewalls. 
This has to be confirmed in more experimental
detail. 
In particular, the comparison  of RB convection in a square 
cell with diagonal (unconfined case)
and  with edge parallel (confined case) wind seems promising.
-- Another aspect is to distinguish the 1-roll 
from the 2-rolls (or more) states. It seems also clear that measuring the large
scale wind velocity should take notice that possibly $U_p \neq U_w$.
Circulation time measurements average over both, while local methods can
distinguish $U_p$ and $U_w$. -- 
Finally, we have
indicated how the global theory of $Nu$ versus $Ra, Pr$ has to be modified to
take care of geometry effects and how the different scalings of bulk
fluctuations and large scale quantities are related. 


{\it  
Acknowledgment:}
We thank X. van Doornum for drawing figure 1. 
The work is part of the research  program of
FOM, which is financially supported by NWO.
It was also supported 
by the European Union (EU) under contract HPRN-CT-2000-00162
and by the German-Israeli Foundation (GIF).




\begin{thebibliography}{10}
\bibitem{bem2}
Note that it is the {\it plate} Reynolds number $Re_p$ which is measured
in  ref.\ \cite{qiu01b} where it is called $Re_\gamma$. It is extracted
from the slope of the (linear) profile of the horizontal velocity 
above the bottom plate.
\end{thebibliography}

\end{document}